\documentclass[twocolumn,
superscriptaddress,
nobibnotes,
amsmath,amssymb,
pre,
]{revtex4-2}

\usepackage[dvipsnames]{xcolor}
\usepackage{graphicx}
\usepackage{siunitx}

\usepackage{multirow}

\definecolor{darkblue}{rgb}{0,0,0.6}
\definecolor{darkred}{rgb}{0.6,0,0}
\definecolor{color_pierre}{rgb}{0.53, 0.0, 0.69}
\usepackage[colorlinks=true,urlcolor=darkblue, citecolor=darkblue,linkcolor=darkred,hyperfootnotes=false]{hyperref}

\newcommand{\dd}{\mathrm{d}}

\newcommand{\ff}{\boldsymbol{f}}
\newcommand{\FF}{\boldsymbol{F}}

\newcommand{\kk}{\boldsymbol{k}}

\newcommand{\RR}{\boldsymbol{R}}

\newcommand{\uu}{\boldsymbol{u}}
\newcommand{\vv}{\boldsymbol{v}}

\newcommand{\xx}{\boldsymbol{x}}
\newcommand{\XX}{\boldsymbol{X}}

\newcommand{\zz}{\boldsymbol{z}}

\newcommand{\eeta}{\boldsymbol{\eta}}

\newcommand{\XXi}{\boldsymbol{\Xi}}

\newcommand{\p}{\partial}

\newcommand{\ind}[1]{_\mathrm{#1}}
\newcommand{\expo}[1]{^{\mathrm{#1}}}
\newcommand{\norm}[1]{\left\lVert#1\right\rVert}
\providecommand{\avg}[1]{\left \langle #1 \right \rangle}
\providecommand{\f}[2]{\frac{#1}{#2}}

\DeclareSIUnit{\molar}{M}
\DeclareSIUnit{\mmol}{\milli \molar}
\DeclareSIUnit{\ums}{\micro\meter\per\second}

\renewcommand{\selectlanguage}[1]{}

\begin{document}
  
  \title{Bath-modes quantitatively capture the nonlinear microrheology of micellar solutions}
  
  \author{Pierre Champagnac}
  \email{pierre.champagnac@espci.psl.eu}
  \affiliation{Gulliver UMR CNRS 7083, ESPCI Paris, PSL Research University, 75005 Paris, France}
  \affiliation{Université Lyon 1, CNRS, ILM, UMR 5306, Villeurbanne, France}
  \author{Clemens Bechinger}
  \affiliation{Fachbereich Physik, Universität Konstanz, 78457 Konstanz, Germany}
  \author{Juliana Caspers}
  \affiliation{Institute for Theoretical Physics, Georg-August Universität Göttingen, 37073 Göttingen, Germany}
  \author{Pierre Illien}
  \affiliation{Sorbonne Université, CNRS, Physicochimie des Électrolytes et Nanosystèmes Interfaciaux (PHENIX), Paris, France}
  \author{Matthias Krüger}
  \affiliation{Institute for Theoretical Physics, Georg-August Universität Göttingen, 37073 Göttingen, Germany}
  \author{Vincent Démery}
  \email{vincent.demery@univ-lyon1.fr}
  \affiliation{Gulliver UMR CNRS 7083, ESPCI Paris, PSL Research University, 75005 Paris, France}
  \affiliation{Université Lyon 1, CNRS, ILM, UMR 5306, Villeurbanne, France}
  \affiliation{Université Lyon, ENSL, CNRS, Laboratoire de Physique, F-69342 Lyon, France}

  \begin{abstract}
    Active microrheology experiments, in which a probe is driven through a complex fluid, often exhibit nonlinear responses that cannot be captured by generalized Langevin equations.
    Models that couple the probe to a Gaussian field reproduce such nonlinear effects qualitatively, but their large number of parameters hinders direct comparison with experiments. Here, we restrict these models to a small number of field modes and demonstrate that this reduced description quantitatively reproduces a broad range of active microrheology experiments in a micellar solution using a single  set of parameters. We further show that the same framework extends naturally to multi-probe systems, such as colloidal dumbbells.
    
  \end{abstract}
  
  \maketitle
  
  
  
  Microrheology aims to determine the rheological properties of a medium via the response of a probe particle moving through it~\cite{furst_Microrheology_2017}.
  If the correlation time of the medium is small compared to   the characteristic time of motion, its dynamics can be faithfully described by a Langevin equation~\cite{einstein_zur_1906, langevin_theorie_1908, perrin_mouvement_1909}.
  If the correlation time of the medium is comparable or large compared to that of probe motion, the dynamics of the probe acquires a memory.
  In case the dynamics of the probe position $\XX(t)$ remains linear, it is then well described  by the Generalized Langevin Equation (GLE)~\cite{mori_transport_1965, kubo_fluctuationdissipation_1966}
  \begin{equation}\label{eq:gle}
    \gamma\dot\XX(t)=-\int_{-\infty}^t\Gamma(t-t')\dot \XX(t')dt'+\eeta(t)+\ff\ind{ext}(t),
  \end{equation}
  where $\gamma$ is an instantaneous Stokes drag coefficient, 
  $\Gamma(t)$ is the memory kernel, $\ff\ind{ext}(t)$ is the external force, and $\eeta(t)$ is a Gaussian noise obeying the fluctuation-dissipation relation.
  The memory kernel $\Gamma(t)$, whose Fourier transform is proportional to the complex modulus of the fluid~\cite{mason_optical_1995, mason_particle_1997}, may be determined from the Mean-Squared Displacement (MSD) of the probe.
  This method has been used to characterize environments as diverse as colloidal suspensions~\cite{mason_optical_1995},
  DNA~\cite{mason_particle_1997}, giant-micelle~\cite{cardinaux_microrheology_2002}, F-actin solutions~\cite{liu_microrheology_2006}, or protein-RNA condensates~\cite{alshareedah_programmable_2021}.

  Dragging the probe with optical~\cite{meyer_laser_2006, brau_passive_2007} or magnetic tweezers~\cite{gosse_magnetic_2002, wilhelm_rotational_2003} gives access to the nonlinear response of the medium. Various protocols have been developed: moving the probe at constant speed~\cite{meyer_laser_2006, wilson_passive_2009, gomez-solano_probing_2014}, in an oscillatory manner~\cite{neckernuss_active_2016, khan_optical_2019}, 
  applying a constant force~\cite{wilking_optically_2008, yang_microrheology_2013}, or measuring recoil by releasing the probe after an imposed displacement~\cite{chapman_nonlinear_2014, gomez-solano_transient_2015, ginot_recoil_2022}.
  Such methods have provided insights on the shear thinning of colloidal suspensions~\cite{meyer_laser_2006, wilson_passive_2009, sriram_active_2010} and micellar solutions~\cite{gomez-solano_probing_2014}, the yielding of colloidal glasses~\cite{habdas_forced_2004}, gelatin~\cite{wilking_optically_2008}, and phospholipid monolayers~\cite{choi_active_2011}, the mechanisms of stress propagation and relaxation in entangled DNA solutions~\cite{chapman_nonlinear_2014, khan_optical_2019, peddireddy_Optical_2022}, DNA-actin mixtures~\cite{fitzpatrick_synergistic_2018} and crosslinked microtubule networks~\cite{yang_microrheology_2013}, and the anisotropy of networks of filaments~\cite{neckernuss_active_2016}.
  However, the GLE~\eqref{eq:gle} is not applicable in many of these cases, and there is no analogous framework to understand and interpret this wealth of experimental results.
  
\begin{figure*}
    \centering
    \includegraphics[width=.9\linewidth]{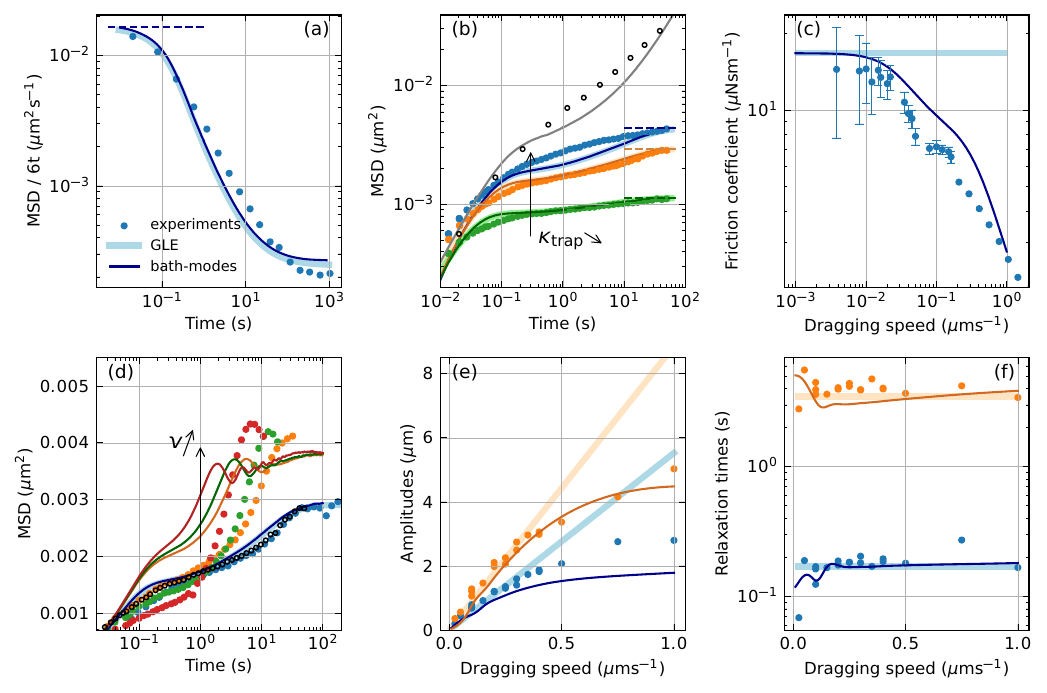}
    \caption{\label{fig:results}
      Results from experiments (dots), GLE (thick light lines) and bath-modes model (thin lines).
      (a) Mean squared displacement (MSD) of the freely diffusing probe rescaled by time.
      (b) MSD of the probe trapped in optical tweezers with stiffness $\qty{1.89}{\micro\N \per\m}$ (blue), $2.81$ (orange) and $7.29$ (green) and no trap (black, same data as (a)).
      (c) Friction coefficient of the probe as a function of the dragging speed.
      (d) MSD of the probe relative to its mean position in a trap with stiffness $\qty{2.81}{\micro\N \per\m}$ moving at $\qty{0}{\ums}$ (blue), 0.12 (orange), 0.19 (green) and 0.37 (red).
      Black circles is the same data as the orange dots in (b). 
      Recoil as a function of the dragging speed, (e) amplitudes $A_i$ and (f) relaxation times $\tau_i$ obtained by fitting the mean recoil, 
      $\langle Z(t)\rangle=\sum_{i=1}^{2}A_i[1-\exp(-t/\tau_i)]$.
    }
  \end{figure*}

  To develop and test an equation of motion capable of capturing the nonlinear dynamics of a probe in a complex medium, it is useful to consider experiments performed on the same system under different driving protocols.
  We focus on silica particles trapped in optical tweezers and immersed in micellar solutions composed of equimolar cetylpyridinium chloride monohydrate (CPyCl) and sodium salicylate (NaSal)~\cite{gomez-solano_probing_2014, gomez-solano_transient_2015, berner_oscillating_2018, muller_properties_2020, jain_two_2021, ginot_recoil_2022, caspers_how_2023, vaidya_observation_2025}~(App.~\ref{app:experiments}).
  We select experiments carried out under identical conditions: probe diameter $\qty{2.73}{\micro\meter}$, concentration $\qty{7}{\mmol}$, and temperature $\qty{25}{\degreeCelsius}$.
  Under these conditions, the solution forms a network of giant wormlike micelles with pronounced viscoelastic properties~\cite{buchanan_high-frequency_2005, berret_linear_1993}.
  We consider three classes of experiments. 
  In passive microrheology, the probe is either freely diffusing~\cite{caspers_how_2023} or confined in a static optical trap~\cite{muller_properties_2020}, and its mean squared displacement $\Delta X^2(t)$ is measured (Fig.~\ref{fig:results}(a,b)). In active microrheology, the probe is driven by a trap moving at constant velocity $v$, allowing one to measure both the friction coefficient (Fig.~\ref{fig:results}(c)) and the fluctuations around the mean position~\cite{berner_oscillating_2018} (Fig.~\ref{fig:results}(d)). Finally, in recoil experiments, the probe is released after being dragged, and its subsequent relaxation is monitored~\cite{ginot_recoil_2022,gomez-solano_transient_2015,caspers_how_2023}. The average recoil is well described by a biexponential form, from which amplitudes and relaxation times can be extracted (Fig.~\ref{fig:results}(e,f)).
  
  Microrheology experiments on micellar solutions have mainly been modeled with bath-particle models.
  In these models, the position $\XX(t)$ of the probe follows an overdamped Langevin equation with friction coefficient $\gamma$ and an interaction with fictitious ``bath particles'' with friction coefficients $\gamma_i$ through a pair potential $V_i(r)$, with $r$ the distance between the probe and a bath particle.
  For experiments in the linear regime, a harmonic potential $V_i(r)=\kappa_i r^2/2$ is appropriate. This allows to integrate out the dynamics of the bath-particles \cite{siegle2010markovian,goychuk2012viscoelastic} to obtain a GLE for the probe with a memory kernel $\Gamma(t)=\sum_i \kappa_i\exp(-\omega_i t)$, where $\omega_i=\kappa_i/\gamma_i$.
  This linear model captures the MSD of a free probe~\cite{caspers_how_2023}, the crossing of an energy barrier~\cite{ginot_barrier_2022} and the recoil dynamics after a time-dependent driving at small velocities~\cite{ginot_recoil_2022, vaidya_observation_2025}.
  Experiments beyond the linear regime have been described with the Stochastic Prandtl-Tomlinson model (SPT), which uses $V_i(r)=-\kappa_i k_i^{-2}\cos(k_i r)$, where $k_i^{-1}$ is a characteristic length~\cite{muller_properties_2020}.
  The SPT describes the shear-thinning behavior and the oscillations of a probe in a moving trap~\cite{jain_two_2021, jain_2021_micro-rheology}.
  These minimal models have thus reached a good agreement with experimental data, but their parameters have been chosen independently for each experiment, leaving open the question of the possibility to describe diverse protocols with a single model.
  
  Environments with finite correlation time and correlation length have traditionally and successfully been modeled in terms of field theory \cite{kardar2007statistical, cha95}. 
  In the context of critical phenomena, the field may represent the concentration of a component of a binary mixture, thereby yielding, e.g., critical exponents found close to the critical point of the mixture, or predictions of the critical Casimir effect~\cite{fisher_phenomenes_1978, hertlein_direct_2008, gambassi_critical_2009, martinez_laserinduced_2023}.
  Inspired by this success, dynamical cases have also been studied \cite{hohenberg_theory_1977}, including the coupling of a probe particle to a scalar Gaussian field $\Phi(\xx,t)$ with overdamped dynamics~\cite{demery_drag_2010, demery_thermal_2011, demery_perturbative_2011, dean_diffusion_2011, demery_driven_2019, basu_dynamics_2022, demery_non-gaussian_2023, venturelli_memory-induced_2023, kruger_stresses_2018, venturelli_universal_2025, pruszczyk_recoil_2025}.
  More recently, coupling a probe to a model A or model B Gaussian field~\cite{hohenberg_theory_1977} has been shown to reproduce qualitatively some experimental observations on viscoelastic fluids, such as shear-thinning~\cite{demery_drag_2010}, recoil dynamics after driving~\cite{pruszczyk_recoil_2025}, and oscillations in a driven trap
  \cite{venturelli_memory-induced_2023}.
  However, models A and B have been developed for near-critical fields, and generalizing the approach to arbitrary fields requires an infinite number of parameters.
  For this reason, these models have never been compared quantitatively to experimental data.
  
\begin{figure}[b]
    \centering
    \includegraphics[width=\columnwidth]{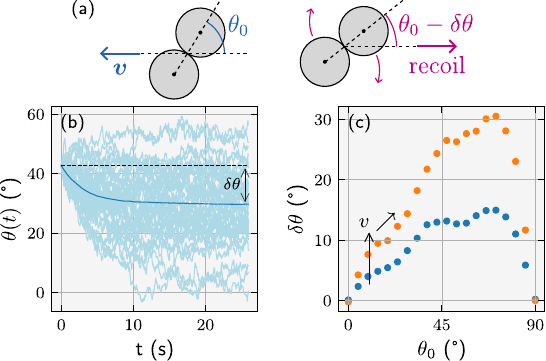}
    \caption{\label{fig:dumbbell}
      Predictions of the bath-modes model for the orientational recoil of a colloidal dumbbell.
      (a) A colloidal dumbbell is pulled with a velocity $\vv$ and an orientation $\theta_0$ with the driving direction.
      When the trap is turned off, the dumbbell recoils and rotates by an angle $\delta\theta$.
      (b) Angular trajectories (light blue) and average over $\num{2000}$ realizations (blue).
      (b) Amplitude of the rotation as a function of the initial angle for  $v=\qty{0.2}{\micro\meter\per\s}$ (blue) and $v=\qty{0.3}{\micro\meter\per\s}$ (orange). 
    }
  \end{figure}

  
  Here, inspired by the bath-particles models, we propose to restrict the above mentioned Gaussian field to a few wavelengths, or ``modes'', where each mode has a finite number of adjustable parameters, namely three. 
  We show that this \emph{bath-modes model} is able to quantitatively reproduce the experimental observations displayed in Fig.~\ref{fig:results} with a single set of parameters.
  The nonlinear character of this model is encoded in the wavelengths of the modes, corresponding to characteristic scales of the perturbed medium; in the large wavelength limit, the coupling to a bath-mode becomes equivalent to the coupling to a bath-particle with a harmonic spring.
  Last, we show that the bath-modes model naturally extends to several probe particles and that it recovers the orientational recoil of a colloidal dumbbell \cite{krishna_kumar_memory-induced_2023}, without any additional ingredient~(Fig.~\ref{fig:dumbbell}).
  
  
  We assume that the probe follows an overdamped Langevin dynamics in the potential $\Phi(\xx,t)$~\cite{demery_perturbative_2011},
  \begin{equation}\label{eq:dyn_probe}
    \gamma\dot\XX(t) = -\nabla\Phi(\XX(t),t)+\eeta(t)+\ff\ind{ext}(t),
  \end{equation}
  where the Gaussian noise $\eeta(t)$ satisfies the fluctuation-dissipation theorem (FDT).
  The restriction to a finite number of wavelengths $\lambda_j=2\pi/k_j$ is imposed by expressing the field with a finite number of Fourier modes,
  \begin{equation}\label{eq:field_fourier}
    \Phi(\xx,t) =- \sum_j\frac{3\omega_j\kappa_j}{4\pi k_j^2}\int_{S_2} e^{ik_j\uu\cdot\xx}\tilde\phi_j(\uu,t)d \uu,
  \end{equation}
  where $\omega_j$ is the relaxation rate of mode $j$, $\kappa_j$ is the strength of its coupling to the probe, and the sum runs over the considered modes.
  $k_j$, $\omega_j$ and $\kappa_j$  are the three mentioned parameters per mode, as for the SPT model.
  To ensure isotropy, all wavevectors $\kk$ with norm $k_i$ are included for each mode, leading to a Fourier transform $\tilde\phi_j(\uu,t)$ defined over the unit sphere $S_2$.
  Last, 
  mode $j$ follows the overdamped Langevin dynamics
  \begin{equation}\label{eq:dyn_field}
    \partial_t\tilde\phi_j(\uu,t) = -\omega_j\tilde\phi_j(\uu,t)+e^{-ik_j\uu\cdot\XX(t)}+\xi_j(\uu, t),
  \end{equation}
  where the Gaussian noise $\xi_j(\uu,t)$ satisfies FDT.
  We thus assume that the noise is given by the FDT even when the system is driven; this hypothesis may not hold and can be relaxed, at the cost of adding additional free parameters to the model.
  The probe and the modes interact reciprocally according to the first term in Eq.~\eqref{eq:dyn_probe} and the second term in Eq.~\eqref{eq:dyn_field}. 
  The bath-modes model is completely defined by Eqs.~(\ref{eq:dyn_probe}-\ref{eq:dyn_field}). 
  
  The mode dynamics~\eqref{eq:dyn_field} can be integrated, leading to a non-Markovian dynamics for the probe~\cite{demery_perturbative_2011}:
  \begin{equation}\label{eq:nlgle}
    \gamma\dot\XX(t)=-\int_{-\infty}^t\FF(\XX(t)-\XX(t'),t-t')dt'+\eeta'(t)+\ff\ind{ext}(t),
  \end{equation}
  where
  \begin{align}
    \FF(\XX,t) &= \XX\sum_j\omega_j\kappa_j e^{-\omega_j t}  g(k_j|\XX|),\label{eq:def_F}\\
    g(y) &=3y^{-3}[\sin(y)-y\cos(y)],\label{eq:def_g}
  \end{align}
  and the noise $\eeta'(t)$ inherits the FDT from the Markovian description of Eqs.~(\ref{eq:dyn_probe}-\ref{eq:dyn_field})~\cite{basu_dynamics_2022}~(App.~\ref{app:model}).
  Equations~(\ref{eq:nlgle}-\ref{eq:def_g}) constitute a complete description of the dynamics, equivalent to Eqs.~(\ref{eq:dyn_probe}-\ref{eq:dyn_field}).
  Equation~\eqref{eq:nlgle} is a non-linear analogue of the GLE~\eqref{eq:gle}, the non-linearity being encoded in the function $g(y)$.
  Note that the memory term can still be written as an integral over the previous times $t'<t$, meaning that the effect of the previous positions is additive, as for the GLE.
  Since $g(0)=1$, the harmonic coupling to a bath particle is recovered in the limit $k_j \rightarrow 0$.
  As a consequence, the parameters $\omega_j$ and $\kappa_j$ can be fitted to experiments in the linear regime, such as the recoil experiments~\cite{ginot_recoil_2022}, before the parameter $k_j$ is adjusted on non-linear experiments.
  

  We choose to use three bath-modes with the parameters given in the first three lines of Table~\ref{tab:parameters}.
  The relaxation rate is given by the ratio of the coupling strength and the friction coefficient, $\omega_j=\kappa_j/\gamma_j$.
  The observables are obtained analytically or numerically by solving Eqs.~(\ref{eq:dyn_probe}, \ref{eq:dyn_field})~(App.~\ref{app:integration}); the results are shown as thin lines in Fig.~\ref{fig:results}.
  Overall, we obtain a good quantitative agreement for all the observables, both for passive and active microrheology experiments.
  In passive microrheology experiments, the model captures the transition from the short-time regime to the long-time regime (Fig.~\ref{fig:results}(a,b)).
  The effective friction coefficient is~(App.~\ref{app:integration}): 
  \begin{equation}\label{eq:friction}
    \gamma\ind{eff} = \gamma+3\sum_j\gamma_j\frac{v_{c,j}^2}{v^2}\left[1-\frac{v_{c,j}}{v}\arctan \left(\frac{v}{v_{c,j}} \right) \right],
  \end{equation}
  where we have introduced the \emph{critical velocity} $v_{c,j} = \omega_j / k_j$ (Table~\ref{tab:parameters}).
  The additional friction due to the mode $j$ is $\gamma_j$ for $v\ll v_{c,j}$ and decays as $v^{-2}$ for $v\gg v_{c,j}$, meaning that the mode does not have time to respond to the presence of the probe.
  The small critical velocity of mode $1$, $v_{c,1}\simeq \qty{0.027}{\micro m\per\second}$, may explain why two bath-particles have been enough to capture the recoil dynamics in Ref.~\cite{ginot_recoil_2022}, which have been performed for velocities above $\qty{0.025}{\micro m\per\second}$~(Fig.~\ref{fig:results}(e,f)).
  On the contrary, the friction experiments of Ref.~\cite{jain_two_2021} explored velocities down to $\qty{0.004}{\micro m\per\second}$, so that matching the friction curve with our model requires the mode 1 (Fig.~\ref{fig:results}(c))~(App.~\ref{app:2v3modes}).
  As the dragging speed increases and becomes comparable to the critical velocities of modes 2 and 3, the bath-modes model captures quantitatively the shear-thinning (Fig.~\ref{fig:results}(c)) and the saturation of the recoil amplitudes (Fig.~\ref{fig:results}(e)).
  The bath-modes model also captures the nonmonotonic behavior of the MSD in a moving trap (Fig.~\ref{fig:results}(e)), with a quantitative agreement for the value of the maximum and the long-time value.
  The model underestimates the time when the maximum is reached by a factor $\approx 3$, 
  but its evolution with the dragging speed is captured correctly.

  \begin{table}[]
    \centering
    \renewcommand{\arraystretch}{1.4}
    \begin{tabular}{lccc}
      \toprule Mode & 1 & 2 & 3 \\ \hline
      \hline
      $\kappa_j/\gamma$ $[\qty{}{\per\s}]$               & $5.0$      & $1.7$    & $3.5$    \\ 
      \hline
      $\gamma_j/\gamma$                                   & $32$     & $30$     & $4.0$      \\ \hline
      $k_i$ $[\qty{}{\per\micro\m}]$                       & $5.79$   & $0.16$   & $0.29$   \\ \hline
      \hline
      $\omega_j = \kappa_j / \gamma_j$ $[\qty{}{\per\s}]$ & $0.16$   & $0.057$  & $0.88$   \\ \hline
      $v_{c,j}=\omega_j/k_j$ $[\qty{}{\micro\m\per\s}]$                & $0.027$  & $0.35$   & $3.0$      \\ \hline
      $\lambda_j = 2\pi / k_j$ $[\qty{}{\micro\m}]$    & $1.1$   & $39$  & $22$  \\ \hline
      $k_j\left[\Delta X^2(\omega_j^{-1})\right]^{1/2}$                          & 0.5    & 0.02  & 0.02  \\ \botrule
    \end{tabular}
    \caption{\label{tab:parameters}
      Parameters (first three lines) and related quantities (last four lines) for the bath-particles and bath-modes models. 
      The bare friction coefficient of the probe is set to  $\gamma=\qty{0.25}{\micro\newton\s\per\meter}$.}
  \end{table}

  We now comment on the parameters used in the bath-modes model.
  The relaxation times of the modes, from $\qtyrange{1}{20}{\second}$, are compatible with macrorheology experiments
  ~\cite{buchanan_high-frequency_2005, berret_linear_1993}.
  The friction coefficients of the modes are $\numrange{4}{30}$ times larger than the bare friction coefficient; as a consequence, the effect of the bath-modes cannot be treated perturbatively, as is usually done to obtain analytical expressions~\cite{demery_perturbative_2011, demery_driven_2019, basu_dynamics_2022, venturelli_memory-induced_2023, demery_non-gaussian_2023, pruszczyk_recoil_2025}.
  This difference also explains why the bare friction coefficient $\gamma$ is not visible at large speed in Fig.~\ref{fig:results}(c).
  The third parameter, the wavelength, is surprisingly large for the modes 2 and 3, respectively $\qtylist{39;22}{\micro m}$, exceeding the size of the probe and the typical length of wormlike micelles, $\qtyrange{0.1}{1}{\micro\meter}$~\cite{buchanan_high-frequency_2005, berret_linear_1993}, in agreement with observations made when fitting the SPT model \cite{jain_two_2021}.
  The origin of these large wavelengths is an open question, which could be addressed by repeating the same analysis at different micellar concentration, temperature, or particle size, and monitoring their evolution.
  
  The prediction of the bath-modes model can be compared to the prediction of the GLE, obtained by setting $k_j=0$ and keeping the parameters $\kappa_j$ and $\gamma_j$ constant (Fig.~\ref{fig:results}, thick light lines).
  As expected, the GLE does not capture the shear-thinning behavior (Fig.~\ref{fig:results}(c)), the saturation of the recoil amplitudes (Fig.~\ref{fig:results}(e)) and neither the non-monotonic MSD in a moving trap (Fig.~\ref{fig:results}(d)).
  However, there is no discernible difference between the results of the bath-modes model and the GLE for the passive microrheology experiments (Fig.~\ref{fig:results}(a,b)).
  That the bath-modes model remains in the linear regime means that the argument of the function $g$ in Eq.~\eqref{eq:def_F}, which involves the displacements during a time $t$, remains small for times $t\lesssim\omega_j^{-1}$.
  The displacement can be estimated from the MSD, leading to the condition $k_j\Delta X^2(\omega_j^{-1})\ll 1$, which is satisfied for our choice of parameters (Table~\ref{tab:parameters}).
  
  
  We have proposed the bath-modes model, which reproduces quantitatively the nonlinear effects appearing in the active microcrheology of a micellar solution.
  As this model couples the probe to spatially extended bath-modes, it naturally extends to several probes, which can be coupled to the same bath-modes.
  To test this hypothesis, we apply our model to the recoil of a colloidal dumbbell in a micellar solution~\cite{krishna_kumar_memory-induced_2023}.
  In these experiments, the dumbbell consists in two silica particles held together by depletion forces.
  Using optical traps, the dumbbell can be driven with an imposed angle $\theta_0$ between the dumbbell and the direction of motion (Fig.~\ref{fig:dumbbell}(a)).
  When the trap is turned off, the dumbbell recoils and rotates by an angle $\delta\theta$, which depends on the inital angle $\theta_0$, being maximal for $\theta_0\approx\qty{45}{\degree}$, and on the dragging velocity.
  Because these experiments have used a different concentration from the experiments discussed above, we restrict ourselves to a qualitative comparison.
  We simulate the coupled dumbbell-field dynamics using the parameters given in Table.~\ref{tab:parameters}~(App.~\ref{app:dumbbell}).
  Evolutions of the dumbbell orientation $\theta(t)$ during its recoil are shown in Fig.~\ref{fig:dumbbell}(b); averaging over many realizations, we obtain the rotation $\delta\theta$, which is plotted as a function of the initial angle and for two driving velocities in Fig.~\ref{fig:dumbbell}(c).
  The bath-modes model thus also captures the rotation of a dumbbell during its recoil without any additional ingredient.
  As a consequence, it can be used not only to describe the nonlinear dynamics of a probe in a viscoelastic medium, but also the coupled dynamics of several probes.
  Its applicability beyond model viscoelastic fluids, for example for critical media, remains to be investigated.

  \begin{acknowledgments}
    We acknowledge insightful discussions with A. Gambassi. 
    P. C. is supported by a Ph. D. grant from ED564 ``Physique en Ile de France”.
    C.B. and M.K. were supported by the Deutsche Forschungsgemeinschaft (DFG), Grant No. SFB 1432–Project ID 425217212, Project C05.
  \end{acknowledgments}
  
  \appendix
  
  \begin{widetext}

  \section{Experiments}\label{app:experiments}
  
  The original publications associated with the data presented in the main text are given in Table~\ref{tab:experiments}.
  
  \begin{table}[h]
    \centering
    \begin{tabular}{|l|l|l|}
      \hline
      Experiment &  Main text fig. & Reference\\ \hline\hline
      MSD of the free probe & Fig.~\ref{fig:results}(a)  & \cite{caspers_how_2023}, Fig.~2(a) \\ \hline
      MSD of the trapped probe at equilibrium & Fig.~\ref{fig:results}(b) & \cite{muller_properties_2020}, inset of Fig.~(10) \\ \hline
      Friction & Fig.~\ref{fig:results}(c) & \cite{jain_two_2021}, Fig.~1 \\ \hline
      MSD of the driven trapped probe & Fig.~\ref{fig:results}(d) & \cite{berner_oscillating_2018}, Figs.~2 and 6\\ \hline
      Recoil & Fig.~\ref{fig:results}(e),(f) & \cite{caspers_how_2023}, unpublished data\\ \hline
    \end{tabular}
    \caption{\label{tab:experiments}
      Original publications associated with the data presented in the main text, together with the figure where the data is shown.}
  \end{table}
  
  \section{Model: from Markovian to non-Markovian}\label{app:model}
  \subsection{Memory term}
  Here we show how to obtain the non-Markovian dynamics for the probe from the Markovian dynamics for the probe and the field.
  We start by integrating the dynamics of the Fourier modes of the field (Eq.~\eqref{eq:dyn_field})
  \begin{equation}
    \tilde{\phi}_j (\uu, t) = \int_{-\infty}^t e^{-\omega_j(t-t^{\prime} )} \left( e^{-ik_j\uu \cdot X(t^{\prime} )} + \xi_j(\uu, t^{\prime} )  \right)dt^{\prime} .
  \end{equation}
  Inserting the result in the equation for the field (Eq.~\eqref{eq:field_fourier}), one obtains 
  \begin{equation}
    \Phi(\xx,t) = - \sum_j \frac{3\omega_j \kappa_j}{4\pi k_j^2} \int_{-\infty}^t e^{-\omega_j(t-t^{\prime} )} \int_{S_2} e^{ik_j\uu\cdot\xx}\left( e^{-ik_j\uu\cdot\XX(t^{\prime} )} +\xi_j(\uu, t^{\prime})\right)d\uu dt^{\prime}  
  \end{equation}
  It can be split into a deterministic part $\Phi_D$ and a noisy part $\Phi_N$:
  \begin{align}
    \Phi_D(\xx, t) &= - \sum_j \frac{3\omega_j \kappa_j}{4\pi k_j^2} \int_{-\infty}^t e^{-\omega_j(t-t^{\prime} )} \int_{S_2} e^{ik_j\uu\cdot(\xx- \XX(t^{\prime} ))} d\uu dt^{\prime}\\
    \Phi_N(\xx,t) &= - \sum_j \frac{3\omega_j \kappa_j}{4\pi k_j^2} \int_{-\infty}^t e^{-\omega_j(t-t^{\prime} )} \int_{S_2} e^{ik_j\uu\cdot\xx}\xi_j(\uu, t^{\prime})d\uu dt^{\prime} 
  \end{align}
  The force on the probe is proportional to the gradient of the field.
  The gradient of $\Phi_D$ involves
  \begin{align}
    \frac{\partial}{\partial z_\mu} \int_{S_2} e^{ik_j \uu \cdot \zz} d\uu &= k_j \frac{\partial}{\partial w_\mu}\int_{S_2} e^{i\uu\cdot\boldsymbol{w}}d\uu = k_j \frac{w_\mu}{w} f^{\prime} (w)
  \end{align}
  where \(\boldsymbol{w} = k_j \zz\) and \(w=\norm{\boldsymbol{w}}\) and we introduced the function 
  \begin{equation}\label{eq:f}
    f(w) = \int_{S_2} e^{i\uu\cdot\boldsymbol{w}}d\uu = \frac{4\pi}{w} \sin (w).
  \end{equation}
  So we have
  \begin{equation}
    f^{\prime} (w) = \frac{4\pi}{w^2}[w\cos (w)-\sin(w)]  ,
  \end{equation}
  and 
  \begin{equation}
    \frac{\partial}{\partial z_\mu} \int_{S_2} e^{ik_j \uu \cdot \zz} d\uu = -k_j^2 z_\mu \frac{4\pi}{3} g(k_{j}z)
  \end{equation}
  with
  \begin{equation}\label{eq:g}
    g(w) = 3w^{-3}\left[ \sin (w) - w\cos (w) \right],
  \end{equation}
  which was introduced in Eq.~\eqref{eq:def_g}.
  So the force coming from $\Phi_D$ reads
  \begin{align}
    - \frac{\partial}{\partial x_\mu}\Phi_D(X(t), t ) 
    & = - \sum_j \int_{-\infty}^t  \omega_{j}\kappa_j e^{-\omega_j(t-t^{\prime} )} \left( X_\mu(t) - X_\mu(t^{\prime} )  \right)  g\left( k_{j}\lVert \XX(t) - \XX(t^{\prime} ) \rVert   \right) dt^{\prime}\\
    & = - \int_{-\infty}^t F_\mu(\XX(t) - \XX(t^{\prime},t-t^\prime) dt^{\prime},
  \end{align}
  where
  \begin{equation}\label{eq:F}
    F_\mu(\xx,t) = \xx \sum_j \omega_j \kappa_j e^{-\omega_j t} g(k_j x ),
  \end{equation}
  which was introduced in Eq.~\eqref{eq:def_F}.
  
  \subsection{Noise}
  
  The force coming from $\Phi_N$ reads
  \begin{align}
    - \frac{\partial}{\partial x_\mu}\Phi_N(\XX(t), t ) = i \sum_j \frac{3\omega_j \kappa_j}{4\pi k_j} \int_{-\infty}^t e^{-\omega_j(t-t^{\prime} )} \int_{S_2}u_\mu e^{ik\uu \cdot \XX (t)} \xi_j(\uu, t^{\prime} ) dt^{\prime} d\uu.
  \end{align}
  The noises \(\eeta \) in  Eq.~\eqref{eq:dyn_probe} and \(\eeta^{\prime} \) in Eq.~\eqref{eq:nlgle} are linked by 
  \begin{equation}
    \eeta^{\prime} (t) = \eeta(t) - \frac{\partial}{\partial x_\mu}\Phi_N(X(t), t ) = \eeta(t) + \XXi(X(t), t).
  \end{equation}
  In order to respect the fluctuation-dissipation relation, the noises \(\eeta\) and \(\xi_j(\uu, t)\) have the correlations
  \begin{align}
    \avg{\eta_\mu(t) \eta_\nu(t^{\prime} )} &= 2 T \gamma\delta(t-t^{\prime}) \delta_{\mu \nu} \\
    \avg{\xi_j(\uu,t)\xi_k(\uu^{\prime} , t^{\prime} )} &= \frac{8\pi k_j^2 T}{3 \omega_j \kappa_j}\delta(t - t')\delta(\uu + \uu^{\prime} ) \delta_{jk}.\label{eq:corr_xi}
  \end{align}  
  We compute the correlations of the noise $\Xi$:
  \begin{align}
    \avg{\Xi_\mu(\xx, t) \Xi_\nu(\xx', t^{\prime}) } &= - \sum_j\sum_k \frac{3\omega_j \kappa_j}{4 \pi k_j}\frac{3 \omega_k \kappa_k}{4\pi k_k}\int_{-\infty}^t \int_{-\infty}^{t^{\prime} } e^{-\omega_j(t-s)} e^{-\omega_k(t^{\prime} -s^{\prime} )} \\ \nonumber
    & \qquad \int_{S_2}\int_{S_2}u_\mu u^{\prime}_\nu e^{ik_j\uu \cdot \xx} e^{ik_k\uu^{\prime} \cdot \xx'} \avg{\xi(\uu, s) \xi(\uu^{\prime} , s^{\prime} )} ds ds^{\prime} d\uu d\uu^{\prime} \\
    &= \frac{8\pi}{3}\left( \frac{3}{4\pi} \right)^2 T \sum_j \omega_{j}\kappa_j e^{-\omega_j(t+t^{\prime} )} \int_{-\infty}^{t^{\prime} } e^{2\omega_j s} \int_{S_2}u_\mu u_\nu e^{ik\uu\cdot(\xx-\xx')} ds d\uu
  \end{align}
  where we used the correlations of the noise $\xi$ given in Eq.~\eqref{eq:corr_xi} and considered the case $t>t^{\prime} $.
  We calculate the last integral over the sphere 
  \begin{align}
    \int_{S_2} u_\mu u_\nu e^{ik\uu\cdot\zz} &= - \frac{\partial}{\partial w_\mu}\frac{\partial}{\partial w_\nu}\int_{S_2}e^{i\uu\cdot \boldsymbol{w}} \\
    &= -\frac{\partial}{\partial w_\nu}\left( \frac{w_\mu}{w} f^{\prime}(w) \right) \\
    &= \delta_{\mu \nu} \frac{f^{\prime} (w)}{w} + \frac{w_\mu w_\nu}{w}\frac{f^{\prime\prime}(w)w - f^{\prime} (w)}{w^2} \\
    &= \frac{4\pi}{3} \left[ \delta_{\mu \nu}g(kz) + k \frac{z_\mu z_\nu}{z} g^{\prime} (kz)\right] 
  \end{align} 
  where we introduced $\boldsymbol{w} = k \zz$ and used the functions \(f\) and \(g\) introduced in Eq.~\eqref{eq:f} and \eqref{eq:g}.  
  After computing the final integral over \(s\) we obtain
  \begin{equation}
    \avg{\Xi_\mu(\xx, t) \Xi_\nu(\xx', t^{\prime}) } = T G_{\mu \nu}(\xx - \xx', t-t^{\prime} ) 
  \end{equation}
  where the tensor $G_{\mu \nu}$ is given by 
  \begin{equation}\label{eq:G}
    G_{\mu \nu}(\xx, t) = - \sum_j \kappa_j e^{-\omega_j|t|}\left[ \delta_{\mu\nu}g(kx) + k_j\frac{x_\mu x_\nu}{x}g^{\prime} (kx) \right].
  \end{equation}
  
  \subsection{Fluctuation-dissipation relation}
  
  We check that the memory kernel \(F_\nu(\xx,t)\)~\eqref{eq:F} and the correlations of the noise~\eqref{eq:G} satify for $t>0$ the fluctuation-dissipation relation~\cite{basu_dynamics_2022} 
  \begin{equation}
    \frac{\partial}{\partial x_\mu}F_\nu(\xx, t) = \partial_t G_{\mu \nu}(\xx, t ).
  \end{equation}

  \section{Model integration}\label{app:integration}
  
  \subsection{Analytical results}
  
  \subsubsection{Friction}
  
  We consider that the probe is moving at constant speed $\vv = v \boldsymbol{e}_x$. 
  We introduce the translated field $\Phi^*(\xx,t)$ = \(\Phi(\xx + \vv t, t)\) .
  The equation for the modes \(\tilde \phi^*_j\)  of $\Phi^*$ is
  \begin{equation}
    \p_t \tilde\phi^*_j(\uu,t) - ik_j\uu\cdot \vv \tilde \phi^*_j(\uu,t) = - \omega_j \tilde \phi^*_j(\uu,t) + 1 + \xi_j(\uu,t).
  \end{equation}
  The mean value of the stationary solution of this equation is 
  \begin{equation}
    \left\langle \tilde\phi\expo{*,st}_j(\uu,t)\right\rangle = \f{1}{\omega - ik_j\uu\cdot \vv}.
  \end{equation}
  Hence the mean value of the stationary translated field is 
  \begin{equation}
    \left\langle\Phi\expo{*,st}(\xx, t)\right\rangle 
    = - \sum_{j}\frac{3\omega_{j}\kappa_j}{4\pi k_j^2} \int_{S_2}   \f{e^{ik_j \uu \cdot \xx}}{\omega - ik_j\uu\cdot \vv} d\uu.
  \end{equation}
  It induces a force on the probe \( \boldsymbol{f} \expo{st} = f\expo{st} \boldsymbol{e}_x\) whose mean is
  \begin{equation}\label{eq:force_stationary}
    \left\langle f\expo{st}\right\rangle = - \boldsymbol{e}_x\cdot\nabla\left\langle\Phi\expo{*,st}(0, t)\right\rangle
    = \sum_j\f{3i\omega_j\kappa_j}{2k_j} \int_0^\pi d\theta~\f{\sin\theta\cos\theta}{\omega_j - ik_{j}v\cos\theta}.
  \end{equation}
  where we have introduced a spherical basis aligned with \(e_x\). 
  Calculating the remaining integral yields
  \begin{equation}
    \left\langle f\expo{st}\right\rangle = -3\sum_j\frac{\omega_j\kappa_j}{vk_j^2}\left[1-\frac{\omega_j}{vk_j}\arctan \left(\frac{vk_j}{\omega_j} \right) \right] = -3\sum_j\gamma_j\frac{v_{c,j}^2}{v}\left[1-\frac{v_{c,j}}{v}\arctan \left(\frac{v}{v_{c,j}} \right) \right].
  \end{equation}
  where we have introduced the \emph{critical velocity} \(v_{c,j} = \omega_j / k_j\).   
  We deduce the effective friction (Eq.~\eqref{eq:friction}):
  \begin{equation}\label{eq:friction_app}
    \gamma\ind{eff} 
    = \gamma-\frac{\left\langle f\expo{st}\right\rangle}{v}
    = \gamma+3\sum_j\gamma_j\frac{v_{c,j}^2}{v^2}\left[1-\frac{v_{c,j}}{v}\arctan \left(\frac{v}{v_{c,j}} \right) \right],
  \end{equation}
  where $\gamma$ is the bare friction coefficient of the probe.
  
  \end{widetext}
  
  \subsubsection{MSD of the trapped and free probe following the GLE}
  
  When the dynamics of the probe is described by the GLE and the probe is either trapped in a harmonic potential or free to diffuse, its MSD can be computed analytically.
  We consider the GLE (Eq.~\eqref{eq:gle})
  \begin{equation}\label{eq:gle_app}
    \gamma\dot\XX(t)=-\int_{-\infty}^t\Gamma(t-t')\dot \XX(t')dt'-\kappa \XX(t)+\eeta^{\prime} (t),
  \end{equation}
  with a memory kernel of the form
  \begin{equation}\label{eq:memory_kernel}
    \Gamma(t) = \sum_{j=1}^N\kappa_j e^{-\omega_j t }.
  \end{equation}
  The correlations of the noise \(\eeta'\) are given by the fluctuation-dissipation relation~\cite{kubo_fluctuationdissipation_1966}:
  \begin{equation}
    \avg{\eta^{\prime} _\mu(t) \eta^{\prime} _\nu(t^{\prime} ) } = T \delta_{\mu \nu} \left[ 2\gamma \delta(t-t^{\prime} ) + \Gamma(|t-t^{\prime} |) \right].
  \end{equation}
  
  Absorbing the bare friction term into the memory kernel, which becomes
  \begin{equation}
    \Gamma(t)=\gamma\delta(t)+\sum_{j=1}^{N}\kappa_j e^{-\omega_j t},
  \end{equation}
  we can rewrite the GLE as 
  \begin{equation}\label{eq:gle2}
    \int_{-\infty}^t\Gamma(t-t')\dot \XX(t')dt'=-\kappa \XX(t)+\eeta^{\prime} (t),
  \end{equation}
  
  We consider the two-time position correlation function
  \begin{equation}
    C(t)=\avg{X(t)X(0)}.
  \end{equation}
  In Laplace space, \(C\) and the memory kernel are related through \cite{muller_properties_2020}
  \begin{equation}
    C(p)=\frac{T}{\kappa}\frac{1}{p+\kappa\Gamma(p)^{-1}}.
  \end{equation}
  The Laplace transform of the memory kernel is
  \begin{equation}
    \Gamma(p)=\gamma+\sum_{i=1}^{N}\frac{\kappa_i}{p+\omega_i}.
  \end{equation}
  
  The Laplace transform of the correlation can be inverted analytically, up to a numerical root finding in the denominator of $C(p)$.
  The mean-squared displacement (MSD) follows as
  \begin{equation}\label{eq:msd_correl}
    \avg{\Delta X^2(t)}=2\left[C(0)-C(t)\right].
  \end{equation}
  
  Taking the limit $\kappa\to 0$ in Eq.~\eqref{eq:msd_correl} provides the MSD of the free probe.

  \subsection{Numerical integration}
  
  The Markovian equations of motion~(\ref{eq:dyn_probe}, \ref{eq:field_fourier}, \ref{eq:dyn_field}) can be integrated numerically by discretizing the sphere so that the field is set by a finite number of coefficients.
  We used the Fibonacci algorithm to discretize the sphere with $N = 200$ points.
  The integral on the sphere in Eq.~\eqref{eq:field_fourier} is replaced by a sum over the vertices,
  \begin{equation}
    \int_{S_2} \dd \uu \longrightarrow \frac{4 \pi}{N}\sum_{\uu \in P},
  \end{equation}
  where $P$ is the polyhedron formed by the $N$ points. 
  Doing so, Eq.~\eqref{eq:field_fourier} becomes
  \begin{equation}\label{eq:dyn_disc_field}
    \Phi(\xx,t) = - \sum_j \frac{3\omega_j\kappa_j}{Nk_j^2} \sum_{\uu \in P} e^{i k_j \uu \cdot \xx} \tilde \phi_j (\uu, t) d \uu.
  \end{equation}
  In Eq.~\eqref{eq:dyn_field}, the deterministic terms are not changed and the correlation of the noise is now given by
  \begin{equation}\label{eq:dyn_disc_field_noise}
    \avg{\xi_j(\uu,t)\xi_k(\uu^{\prime} , t^{\prime} )} = \frac{2 N k^2 T}{3 \omega \kappa}\delta(t - t')\delta_{\uu,-\uu'} \delta_{jk}
  \end{equation}
  The stochastic dynamics defined by Eqs.~(\ref{eq:dyn_probe}, \ref{eq:dyn_disc_field}, \ref{eq:dyn_field}, \ref{eq:dyn_disc_field_noise}) is integrated numerically using the stochastic Heun scheme~\cite{kloeden1992NumericalSolution}.
  
  \section{Three vs. two bath modes}\label{app:2v3modes}

\begin{figure*}
    \centering
    \includegraphics[width=.9\linewidth]{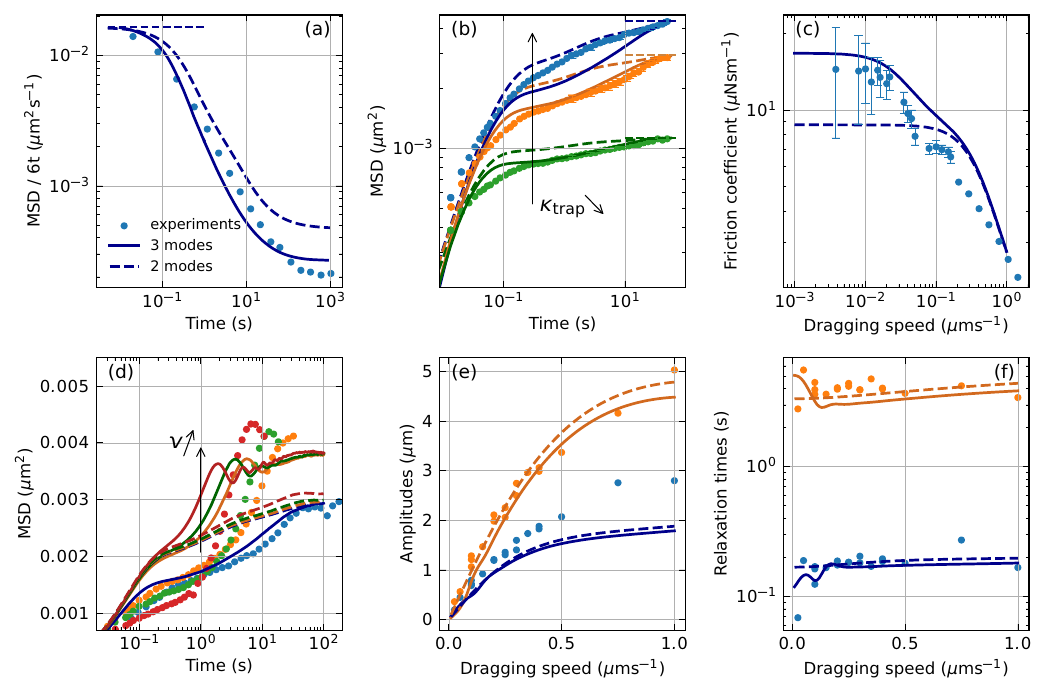}
    \caption{\label{fig:2v3modes}
      Results from experiments (dots), and bath-modes model with 3 modes (solid lines) and 2 modes (dashed lines).
      (a) Mean squared displacement (MSD) of the freely diffusing probe rescaled by time.
      (b) MSD of the probe trapped in optical tweezers with stiffness $\qty{1.89}{\micro\N \per\m}$ (blue), $2.81$ (orange) and $7.29$ (green) and no trap (black, same data as (a)).
      (c) Friction coefficient of the probe as a function of the dragging speed.
      (d) MSD of the probe relative to its mean position in a trap with stiffness $\qty{2.81}{\micro\N \per\m}$ moving at $\qty{0}{\ums}$ (blue), 0.12 (orange), 0.19 (green) and 0.37 (red).
      Black circles is the same data as the orange dots in (b). 
      Recoil as a function of the dragging speed, (e) amplitudes $A_i$ and (f) relaxation times $\tau_i$ obtained by fitting the mean recoil, 
      $\langle Z(t)\rangle=\sum_{i=1}^{2}A_i[1-\exp(-t/\tau_i)]$.
    }
  \end{figure*}

  In Fig.~\ref{fig:2v3modes}, we compare the experimental results (dots) with the predictions of the bath-modes model with the 3 modes in Table~\ref{tab:parameters} (solid lines), and with only the modes 2 and 3 of Table~\ref{tab:parameters} (dashed lines).
  
  Removing the mode 1 does not affect the predictions for the recoil experiments (Fig.~\ref{fig:2v3modes}(e, f)), consistently with the fact that two bath-particles with harmonic coupling to the probe have been enough to capture the recoil dynamics in Ref.~\cite{ginot_recoil_2022}.
  The MSD in a static trap is weakly affected as well (Fig.~\ref{fig:2v3modes}(b)).
  The MSD of a free probe predicted with 2 modes is in quantitative agreement with the experiment at small and intermediate times but overestimates the long time diffusion coefficient (Fig.~\ref{fig:2v3modes}(a)).
  
  The effect of removing the mode 1 is strongest for the friction coefficient, where the short time plateau is strongly underestimated (Fig.~\ref{fig:2v3modes}(c)) and for the MSD in a moving trap, where the oscillations are no longer predicted and the long time limit is underestimated (Fig.~\ref{fig:2v3modes}(d)).

  \section{Dumbbell}\label{app:dumbbell}
  
  \subsection{Model}
  
  We aim to model the drag and release experiment of a dumbbell, a pair of beads rigidly bound together, in a micellar solution \cite{krishna_kumar_memory-induced_2023}. 
  The dumbbell consists of two beads at positions $\XX_\nu(t)$, $\nu\in\{-1,1\}$, held at a fixed center-to-center distance $l = \qty{2.73}{\micro\meter}$ (equal to one silica bead diameter). 
  The motion of the beads is restricted to a plane.
  The positions of the beads are related to the position of the center of mass $\RR(t)$ and the orientation of the dumbbell $\theta(t)$ through
  \begin{equation}
    \XX_\nu(t) = \RR(t) + \frac{\nu l}{2} \begin{pmatrix}
      \cos\theta(t)\\\sin\theta(t)
    \end{pmatrix}.
  \end{equation}
  
  \subsection{Equations of motion}
  
  The dynamics of the field is given by Eq.~\eqref{eq:dyn_field}, where the effects of the two beads add up:
  \begin{equation}\label{eq:dumbbell_field_dynamics}
    \partial_t\tilde{\phi}_j(\uu,t) = -\omega_j\,\tilde{\phi}_j(\uu,t) 
    + \sum_\nu e^{-ik_j\uu\cdot\XX_\nu(t)}
    + \xi_j(\uu,t).
  \end{equation}
  
  In order to determine the dynamics of the dumbbell, we identify the forces due to the coupling to the field acting on the beads in Eq.~\eqref{eq:dyn_probe}:
  \begin{equation}
    \ff_{\nu}(t)=-\nabla\Phi\!\left(\XX_{\nu}(t),t\right).
  \end{equation}
  The position of the center of mass responds to the total force on the dumbbell,
  \begin{equation}\label{eq:dumbbell_cm_dynamics}
    \Gamma\dot\RR(t)=\sum_\nu\ff_\nu(t)+\eeta_\mathrm{CM}(t),
  \end{equation}
  where $\Gamma$ is the friction coefficient of the center of mass, which is the sum of the individual bead friction coefficients $\gamma$, $\Gamma=2\gamma$, and $\eeta_\mathrm{CM}(t)$ is a Gaussian white noise satisfying the fluctuation-dissipation relation.
  
  The orientation of the dumbbell evolves according to the torque
  \begin{equation}
    \tau(t)=\sum_\nu \left[\frac{\nu l}{2} \begin{pmatrix}
      \cos\theta(t)\\\sin\theta(t)
    \end{pmatrix} \times \ff_\nu(t)\right]\cdot\boldsymbol{e}_z,
  \end{equation}
  where $\boldsymbol{e}_z$ is the unit vector orthogonal to the plane of motion.
  The orientation thus follows
  \begin{equation}\label{eq:dumbbell_angular_dynamics}
    \gamma_\mathrm{rot}\,\dot{\theta}(t) = \tau(t) + \eta_\mathrm{rot}(t),
  \end{equation}
  where the rotational friction coefficient 
  $\gamma_\mathrm{rot}$ is approximated by that of a prolate ellipsoid with semi-axes 
  $l/2$, $l/2$, and $l$, $\gamma_\mathrm{rot} = l^2\gamma$, and $\eeta_\mathrm{rot}(t)$ is a Gaussian white noise satisfying the fluctuation-dissipation relation.
  
  \subsection{Protocol}
  
  As for the experiment, the numerical simulation consists in two phases.
  During the driving phase, the dumbbell is driven at constant velocity $-v$ along the $x$-axis while held at a fixed angle $\theta_0$ with respect to the driving direction.
  The dynamics of the field~(Eq.~\eqref{eq:dumbbell_field_dynamics}) is integrated numerically until a stationnary state is reached.
  
  During the recoil phase, the dynamics of the field (Eq.~\eqref{eq:dumbbell_field_dynamics}) and of the center of mass (Eq.~\eqref{eq:dumbbell_cm_dynamics}) and orientation (Eq.~\eqref{eq:dumbbell_angular_dynamics}) of the dumbbell are integrated numerically without external force.
  
  The numerical parameters are the ones given in Table~\ref{tab:parameters}.

  \bibliography{Bath_models}

\end{document}